\documentclass[12pt]{article}

\usepackage{amssymb,amsfonts,amsthm}
\usepackage{amsmath}
\usepackage{bm}             
\usepackage[mathscr]{eucal}
\usepackage{cancel}
\usepackage{wasysym}

\def\be{\begin{equation}}
\def\ee{\end{equation}}
\def\bea{\begin{eqnarray}}
\def\eea{\end{eqnarray}}

\def\bdelta{\mbox{\boldmath $\delta$}}

\def\hsp5{\hspace{5mm}}

\theoremstyle{remark}

\newcommand{\sfrac}[2]{{\textstyle{#1\over#2}}}

\title{\sc Second order cosmological perturbations: new conserved quantities
and the general solution at super-horizon scale}

\begin{document}

\author{ \\
{\Large\sc Claes Uggla}\thanks{Electronic address:
{\tt claes.uggla@kau.se}} \\[1ex]
Department of Physics, \\
Karlstad University, S-651 88 Karlstad, Sweden
\and \\
{\Large\sc John Wainwright}\thanks{Electronic address:
{\tt jwainwri@uwaterloo.ca}} \\[1ex]
Department of Applied Mathematics, \\
University of Waterloo,Waterloo, ON, N2L 3G1, Canada \\[2ex] }

\date{}
\maketitle
\bigskip

\begin{abstract}

The study of long wavelength scalar perturbations, in particular the existence
of conserved quantities when the perturbations are adiabatic,
plays an important role in e.g. inflationary cosmology. In this paper
we present some new conserved quantities at second order and
relate them to the curvature perturbation in the uniform density gauge,
$\zeta$, and the comoving curvature perturbation, ${\cal R}$. We also,
for the first time, derive the general solution of the perturbed
Einstein equations at second order, which thereby contains both growing and
decaying modes, for adiabatic long wavelength perturbations for a stress-energy
tensor with zero anisotropic stresses and zero heat flux. The derivation uses
the total matter gauge, but results are subsequently translated to the
uniform curvature and Poisson (longitudinal, zero shear) gauges.

\end{abstract}

\section{Introduction}

In this paper we consider first and second order scalar perturbations of
Friedmann-Lema\^{i}tre (FL) universes subject to the following assumptions:
i) the spatial background is flat;
ii) the stress-energy tensor can be written in the form
\begin{equation}  \label{Tab}
T^a\!_b = \left(\rho + p\right)\!u^a u_b + p\delta^a\!_b, \qquad u^au_a=-1,
\end{equation}
thereby describing perfect fluids and scalar fields;
iii) the linear perturbation is purely scalar.
This paper, which deals with perturbations on super-horizon scales,
relies heavily on two previous papers which we shall refer to
as UW1~\cite{uggwai19a} (a unified and simplified formulation of
change of gauge formulas at second order)
and UW2~\cite{uggwai18} (five ready-to-use systems of governing equations
for second order perturbations).

Two gauge invariants that are conserved for adiabatic long wavelength
perturbations at first and second order play an important role in e.g.
inflationary cosmology, namely, the curvature perturbation in the
uniform density gauge, labelled $\zeta$, and the curvature perturbation
in the total matter gauge, labelled $ {\cal R}$, also often referred to
as the comoving curvature perturbation. We briefly discuss the history of
these conserved quantities and give references at the beginning of
section~\ref{cons}. In this paper we present some new conserved quantities
that in contrast are associated with the uniform curvature gauge.
In particular, writing the perturbed Einstein equations in the
super-horizon regime in that gauge suggests consideration of a gauge invariant
which we denote by $\chi_{\mathrm c}$, defined in terms of $\phi_{\mathrm c}$,
the purely temporal metric perturbation (see the next section) in the uniform
curvature gauge according to:
\begin{equation} \label{def_chi}
{}^{(1)}\!\chi_{\mathrm c} = (1+q)^{-1}{}^{(1)}\!\phi_{\mathrm c}, \qquad
{}^{(2)}\!\chi_{\mathrm c} =(1+q)^{-1} \left({}^{(2)}\!\phi_{\mathrm c} -
4{}^{(1)}\!\phi_{\mathrm c}^2\right),
\end{equation}
where $q$ is the background deceleration parameter. At first order one of
the perturbed Einstein equations shows that
$\chi_{\mathrm c}$ is a conserved quantity,
while the two constraint equations relate the density and
velocity perturbations algebraically to $\chi_{\mathrm c}$,
thereby providing two more conserved quantities. In addition these
equations show that $\chi_{\mathrm c}$ in fact coincides with ${\cal R}$
at first order.
%
%

Unlike ${\cal R}$ and $\zeta$ these new quantities are not conserved at
second order. However, new conserved quantities can be constructed at
second order by adding a certain quadratic source term to the perturbations.
In particular, we use ``source compensated'' second order perturbation
variables of the form that we introduced in an earlier paper
UW1~\cite{uggwai19a} in order to simplify the change of gauge formulas,
which, moreover, are used to relate the new conserved quantities to $\zeta$
and ${\cal R}$ at second order.

We then derive the general solution of the governing equations for adiabatic
long wavelength perturbations at first and second order
subject to the restrictions i)-iii) above. We have found that the
governing equations \emph{in the total matter gauge} are particularly
simple to solve, even when keeping both modes (growing and decaying).
The time dependence of the growing mode of
the first order perturbations is governed
by a function $g(a)$, defined by\footnote{The integral
in~\eqref{def_g_simple}
has a lengthy history in linear perturbation theory,
but a standard symbol for it has not been introduced.
Because of the importance of the function $g(a)$ in
cosmological perturbation theory, we digress in section~\ref{function_g}
to describe some of its history and properties.}
\begin{equation} \label{def_g_simple}
g(a) = 1 - \frac{\cal H}{a^2} \int_0^a \frac{\bar a}{{\cal H}(\bar a)}d{\bar a},
\end{equation}
where ${\cal H}=aH$, with $H$ being the background
Hubble parameter, and $a$ is
the background scale factor. A main result of this paper is to show that
\emph{the simple form of the first order solution
in the total matter gauge extends to
second order}. The
conserved quantities referred to above
emerge naturally in the solution process as temporal
constants of integration (arbitrary spatial functions).
Because of the central role played by the function $g(a)$ we shall refer to it
as the \emph{perturbation evolution function}.

The outline of the paper is as follows. In section~\ref{hat_var},
we introduce the notation for the metric and matter variables
from UW1~\cite{uggwai19a} and UW2~\cite{uggwai18}.
In section~\ref{sec:goveq} we specialize the governing equations to second
order given in UW2~\cite{uggwai18} to long wavelength perturbations.
In section~\ref{cons} we derive the new conserved quantities at second order
and relate them to the previously known ones.
In section~\ref{solutions} we derive the general solution of the
governing equations up to second order in the total matter
gauge, and subsequently transform the results to the uniform curvature and
Poisson gauges by means of gauge transformation rules, followed by
some illustrative applications in Section~\ref{applications}.
In section~\ref{function_g} we give a brief discussion of the history
and properties of the perturbation evolution function $g(a)$.
Section~\ref{discussion} contains the concluding remarks. In the appendices
we give  some background material from
UW1~\cite{uggwai19a} and UW2~\cite{uggwai18}.

\section{Perturbation variables \label{hat_var}}

We describe scalar perturbations of a flat Robertson-Walker geometry by
writing the metric in the form\footnote{The scalar perturbations at
first order will generate vector and
tensor perturbations at second order, but we do not
give these perturbation variables since we will not consider these
modes in this paper.}
\begin{equation}  \label{pert.metric}
ds^2 = a^2\left(-(1+2\phi) d\eta^2 +  {\bf D}_i B\,d\eta dx^i +
(1-2\psi)\delta_{ij} dx^i dx^j \right),
\end{equation}
where $\eta$ is conformal time, the $x^i$ are Cartesian background coordinates
and ${\bf D}_i = \partial/\partial x^i$. The background geometry is described by the
scale factor $a$ which determines the conformal Hubble scalar and the
deceleration parameter according to ${\cal H}=a'/a$ and $q=-{\cal H}'/{\cal H},$
where $'$ denotes differentiation with respect to $\eta$.
By expanding the functions $\phi, B, \psi$ in
a perturbation series\footnote
{A perturbation series for a variable $f$ is a Taylor series in a perturbation
parameter $\epsilon$, of the form
$f= f_0 +\epsilon\,{}^{(1)}\!f + \sfrac12 \epsilon^2\,{}^{(2)}\!f + \dots.$}
we obtain the following metric perturbations up to second order:
 \begin{equation}
 {}^{(r)}\!\phi, {\cal H}{}^{(r)}\!B, {}^{(r)}\!\psi,\quad r=1,2,
 \end{equation}
where the factor of
${\cal H}$ ensures that the $B$-perturbation is dimensionless
(see UW1~\cite{uggwai19a} and UW2~\cite{uggwai18}).

The background matter content is described by
the matter density and pressure, $\rho_0$ and $p_0$,
with associated scalars $w=p_0/\rho_0$ and $c_s^2=p_0'/\rho_0'$.
We will need the fact that the background Einstein equations relate
$w$ and $q$ according to
\begin{equation}  \label{w,q.relation}
 3(1+w)=2(1+q),
 \end{equation}
 (see UW2~\cite{uggwai18}).
The scalar matter perturbations are defined by expanding
$\rho,p,V$ in a perturbation series, where the scalar
velocity potential $V$ is defined in terms of the spatial
\emph{covariant} 4-velocity components
 by $u_i = a{\bf D}_i V$. As in UW2~\cite{uggwai18}, section II.C,
 we scale the density perturbations
 according to ${}^{(r)}\!{\bdelta} = {{}^{(r)}\!\rho}/(\rho_0 + p_0),
 \,r=1,2,$ and replace
 the pressure perturbations ${}^{(r)}\!p$ by the non-adiabatic pressure
 perturbations ${}^{(r)}\!\Gamma,\, r=1,2$, which are defined to be gauge
 invariants with the property that they are zero
 for adiabatic perturbations.\footnote
 {See UW2~\cite{uggwai18}, section II.C, for the definitions. The details are
 not needed in this paper: since we are working exclusively
 with adiabatic perturbations, the terms in
 the perturbation equations that involve
 ${}^{(r)}\!\Gamma,\, r=1,2$, will be set to zero.} Thus the scalar matter perturbations are described up to second order by
the variables
\begin{equation}
{\cal H}{}^{(r)}\!V, \,{}^{(r)}\!\bdelta,\,  {}^{(r)}\!\Gamma,\qquad r=1,2,
\end{equation}
where the factor of ${\cal H}$ ensures that the $V$-perturbation is
dimensionless.
In keeping with this approach we also use the background $e$-fold time
variable $N=\ln (a/a_0)$, where $a_0$ denotes
some reference epoch. For changing to conformal time, note that
$\partial_{\eta} = {\cal H}\partial_{N}$,
$\partial_{\eta}^2 = {\cal H}^2(\partial_{N}^2 - q\partial_{N})$.

In this paper we will show that when studying perturbations
on super-horizon scale significant simplifications
arise when one makes use of the so-called
source-compensated second order perturbation variables,
labelled by a hat on the kernel, that we introduced in our earlier
paper UW1~\cite{uggwai19a}:
\begin{subequations} \label{hatted_var}
\begin{align}
{}^{(2)}\!{\hat{\phi}} &= {}^{(2)}\!{\phi} -2 {}^{(1)}\!{\phi}^2, \\
{}^{(2)}\!{\hat{\psi}} &= {}^{(2)}\!{\psi} +2 {}^{(1)}\!{\psi}^2, \label{hat1} \\
{\cal H} {}^{(2)}\!{\hat{B}} &= {\cal H} {}^{(2)}\!{B} +(1+q)({\cal H} {}^{(1)}\!{B})^2, \\
{\cal H} {}^{(2)}\!{\hat{V}} &= {\cal H} {}^{(2)}\!{V} +(1+q)({\cal H} {}^{(1)}\!{V})^2, \\
{}^{(2)}\!{\hat{\bdelta}} &= {}^{(2)}\!{\bdelta} -(1+c_s^2) {}^{(1)}\!{\bdelta}^2.
\end{align}
\end{subequations}

As regards gauge freedom, in using the line-element~\eqref{pert.metric}
we have fixed the spatial gauge following UW1~\cite{uggwai19a}.
The remaining gauge freedom is the choice of temporal
gauge which we can fix to second order
by setting to zero the first and second perturbations of one the variables
$\psi$, $B$, $V$, ${\bdelta}$. We use the following terminology and subscripts
to label the gauges as in UW1~\cite{uggwai19a}:
\begin{itemize}
\item[i)] $B = 0$, Poisson (longitudinal, zero shear) gauge,
subscript $_{\mathrm p}$, e.g., $\psi_{\mathrm p}$,
\item[ii)] $\psi = 0$, uniform curvature (flat) gauge, subscript $_{\mathrm c}$, e.g., $B_{\mathrm c}$,
\item[iii)] $V = 0$, total matter gauge, subscript $_{\mathrm v}$, e.g., $\psi_{\mathrm v}$,
\item[iv)] ${\bdelta} = 0$, uniform density gauge, subscript $_{\rho}$, e.g., $\psi_{\rho}$.
\end{itemize}
We note in passing that on super-horizon scales the uniform density gauge is equivalent
to the total matter gauge to second order (see appendix~\ref{app:changegauge}).

\section{Governing equations in the super-horizon regime \label{sec:goveq}}

In this section we obtain the governing equations for
perturbations at second order that
we need in this paper by specializing the general
equations in UW2~\cite{uggwai18} to super-horizon scales.
This is accomplished by dropping terms of degree two and higher
in the dimensionless spatial differential operator ${\cal H}^{-1}{\bf D}_i$.
We will use the symbol $\approx$
 to indicate that two expressions are equal once such terms have been dropped.
\subsection{The energy conservation equation}

The perturbed energy conservation equation on super-horizon scales plays a central role
in deriving conserved quantities in cosmological perturbation theory.
By specializing the general perturbed energy conservation equation in
UW2~\cite{uggwai18} (see section IV) to super-horizon scales we obtain
 at first order
\begin{equation} \label{cons_energy.1}
\partial_N({}^{(1)}\!\bdelta - 3{}^{(1)}\!\psi) + 3{}^{(1)}\!\Gamma \approx 0,
\end{equation}
and at second order,
\begin{subequations}\label{cons_energy.2}
\begin{equation} \label{cons_energy.2.1}
\partial_N({}^{(2)}\!\bdelta - 3{}^{(2)}\!\psi) + 3{}^{(2)}\!\Gamma + {\mathbb E} \approx 0,
\end{equation}
where the source term is given by
\begin{equation}
{\mathbb E} \approx - \partial_N(6\psi^2 + (1+c_s^2)\bdelta^2 +
2\bdelta\Gamma) - 6\Gamma^2,
\end{equation}
\end{subequations}
after simplifying them using the first order equation. Here and elsewhere, in the interests of
notational simplicity, we will drop the superscript ${}^{(1)}$
on first order perturbations, when there is no risk of confusion. This applies
in particular to expressions for source terms.
On introducing the hatted variables given in~\eqref{hatted_var},
equation~\eqref{cons_energy.2.1} takes the simpler
form
\begin{equation} \label{cons_energy.2.hat}
\partial_N({}^{(2)}\!\hat{\bdelta} - 3{}^{(2)}\!\hat{\psi})
+ 3({}^{(2)}\!{\Gamma} - 2{\Gamma}^2)
- 2\partial_N(\bdelta\Gamma) \approx 0.
\end{equation}
When specialized to the uniform density gauge (${}^{(r)}\!\bdelta=0,\,r=1,2$) we obtain
\begin{equation}
\partial_N{}^{(1)}\!{\psi}_{\rho} \approx
{}^{(1)}\!{\Gamma}, \qquad
\partial_N{}^{(2)}\!\hat{\psi}_{\rho} \approx
{}^{(2)}\!{\Gamma} - 2{\Gamma}^2.
\end{equation}
These equations are a concise form of well known equations in the literature
(see Malik and Wands (2004)~\cite{malwan04}, equations (5.34) and (5.35),
and Bartolo \emph{et al} (2004)~\cite{baretal04a}, equations (144) and (147)).

\subsection{Governing equations in the total matter gauge  \label{subsubsec:totmat}}

We use the governing equations as given in
UW2~\cite{uggwai18} (see section V.C.1).
At first order we have
\begin{subequations}  \label{totmat_1_new}
\begin{align}
{}^{(1)}\!\phi_{\mathrm v} &= -c_s^2 {}^{(1)}\!\bdelta_{\mathrm v}
-{}^{(1)}\!\Gamma, \label{totmat_1.1} \\
\partial_N {}^{(1)}\!{\psi}_{\mathrm v}
 &= -{}^{(1)}\!\phi_{\mathrm v}, \label{totmat_1.2new}   \\
\partial_N(a^2\,{}^{(1)}\!B_{\mathrm v} ) &=
a^{2}{\cal H}^{-1}\left({}^{(1)}\!{\psi}_{\mathrm v} -{}^{(1)}\!\phi_{\mathrm v}\right) ,  \label{totmat_1.3_new}
\end{align}
\end{subequations}
while the second order equations can be written as
\begin{subequations}\label{totmat_2_new}
\begin{align}
{}^{(2)}\!\phi_{\mathrm v} &= -c_s^2 {}^{(2)}\!\bdelta_{\mathrm v}  -
{}^{(2)}\!\Gamma - {\mathbb M}_{\mathrm v}, \label{phi_v2} \\
\partial_N {}^{(2)}\!{\psi}_{\mathrm v}  &=
-{}^{(2)}\!\phi_{\mathrm v} +
\sfrac12{\mathbb G}^q_{\mathrm v},  \label{totmat_2.2}  \\
\partial_N(a^2\,{}^{(2)}\!B_{\mathrm v} ) &=
a^{2}{\cal H}^{-1}\left({}^{(2)}\!{\psi}_{\mathrm v}
-{}^{(2)}\!\phi_{\mathrm v}
+ {\mathbb G}^{\pi}_{\mathrm v}\right),   \label{totmat_2.3_new}
\end{align}
\end{subequations}
where the source terms  can be obtained from
UW2~\cite{uggwai18}, the Einstein terms ${\mathbb G}^q_{\mathrm v}$
and ${\mathbb G}^{\pi}_{\mathrm v}$ from Appendix A1
and ${\mathbb M}_{\mathrm v}$ from Appendix A3.
In the super-horizon regime equation~\eqref{delta_v_zero} below gives
${}^{(r)}\!{\bdelta}_{\mathrm v}\approx 0$, $r=1,2$, which
for adiabatic perturbations, ${}^{(r)}\!\Gamma=0,\,r=1,2,$ implies
${}^{(1)}\!\phi_{\mathrm v} \approx 0$ and
$\partial_N {}^{(1)}\!{\psi}_{\mathrm v}\approx 0$ by~\eqref{totmat_1_new}.
With these restrictions the source terms reduce to
\begin{equation}  \label{source_ein_v2}
{\mathbb M}_{\mathrm v}\approx0, \qquad
{\mathbb G}^{q}_{\mathrm v}\approx0, \qquad
{\mathbb G}^{\pi}_{\mathrm v} \approx 2\psi_{\mathrm v}^2 -
2{\mathbb D}_0(\psi_{\mathrm v}).
\end{equation}
The differential operator ${\mathbb D}_0$
in~\eqref{source_ein_v2}, which we refer to as the
GR spatial operator, is defined by\footnote
{The \emph{GR spatial operator} ${\mathbb D}_0(C)$
plays a central role in determining the spatial dependence of
second order perturbations at super-horizon scale,
a general relativistic phenomenon
(see, for example, Bartolo \emph{et al} (2006)~\cite{baretal06}). Usually it is written out
in full which makes the source terms look unnecessarily complicated.
See Appendix B of our paper
UW1~\cite{uggwai19a} for some history and properties of ${\mathbb D}_0(C)$.}
\begin{equation} \label{D_0}
{\mathbb D}_0(C) := {\cal S}^{ij}({\bf D}_iC)({\bf D}_jC).
\end{equation}
The scalar mode extraction operator ${\cal S}^{ij}$
is given by
${\cal S}^{ij} = \sfrac32({\bf D}^{-2})^2{\bf D}^{ij}$,
where ${\bf D}_{ij} := {\bf D}_{(i}{\bf D}_{j)} - \sfrac13 \gamma_{ij}{\bf D}^2$
and ${\bf D}^{-2}$ is the inverse Laplacian operator.
The operator ${\mathbb D}_0$ satisfies the identity
\begin{equation}
{\cal S}^{ij}[C{\bf D}_{ij}C]=
\sfrac12 C^2 -{\mathbb D}_0(C),
\end{equation}
which is needed in simplifying the source terms to get~\eqref{source_ein_v2}.

With~\eqref{source_ein_v2} it follows that
 for long wavelength adiabatic perturbations
equations~\eqref{totmat_1_new} and~\eqref{totmat_2_new}
reduce to the very simple form:
\begin{subequations} \label{totmat_sh_adiabatic}
\begin{equation} \label{comov2_sh}
\partial_N{}^{(r)}\!{\psi}_{\mathrm v} \approx 0, \qquad
{}^{(r)}\!\phi_{\mathrm v}\approx0, \qquad
{}^{(r)}\!{\bdelta}_{\mathrm v}\approx 0, \qquad r=1,2,
\end{equation}
with $B_{\mathrm v}$ determined by
\begin{align}
\partial_N(a^2\,{}^{(1)}\!B_{\mathrm v} ) &\approx
a^{2}{\cal H}^{-1}{}^{(1)}\!{\psi}_{\mathrm v},  \\
\partial_N(a^2\,{}^{(2)}\!B_{\mathrm v} ) &\approx
a^{2}{\cal H}^{-1}\left({}^{(2)}\!\hat{\psi}_{\mathrm v}
- 2{\mathbb D}_0({}^{(1)}\!\psi_{\mathrm v})\right). \label{B2.evol}
\end{align}
\end{subequations}
For convenience we have incorporated part of the source term in~\eqref{B2.evol}
into ${}^{(2)}\!\psi_{\mathrm v}$ to give ${}^{(2)}\!\hat{\psi}_{\mathrm v}$.

\subsection{Governing equations in the uniform curvature gauge}  \label{subsubsec:supuni}

In this section we make use of the governing equations
in the uniform curvature gauge, given in
UW2~\cite{uggwai18} (see section V.B.1).
In appendix~\ref{evol_ucg} we
specialize these equations to the super-horizon regime
(see equations~\eqref{ucg_gov1} and~\eqref{ucg_gov2}).
The form of these equations suggest that
we introduce the
new variable $\chi_{\mathrm c}$ defined by~\eqref{def_chi},
which at first order leads to
\begin{equation}\label{ucg_gov1sup}
\partial_N {}^{(1)}\!{\chi_{\mathrm c}} \approx 0, \qquad
{\cal H}{}^{(1)}\!{V}_{\mathrm c} =  -{}^{(1)}\!{\chi_{\mathrm c}},\qquad
{}^{(1)}\!{\bdelta}_{\mathrm c} \approx -3 {}^{(1)}\!{\chi_{\mathrm c}},
\end{equation}
After using these first order equations to write
the source terms~\eqref{source.T_c} in terms of $\chi_{\mathrm c}$
the equations at second order assume the form
\begin{subequations} \label{ucg_gov2sub}
\begin{align}
\partial_N {}^{(2)}\!{\chi_{\mathrm c}} &\approx
\partial_N\left[ - 3(1 + c_s^2)\chi_{\mathrm c}^2 \right],  \label{ucg_gov2.1sub}  \\
{\cal H}{}^{(2)}\!{V}_{\mathrm c} &\approx  - {}^{(2)}\!{\chi_{\mathrm c}}
 - [3(1+c_s^2)+ (1+q)]\chi_{\mathrm c}^2,  \label{ucg_gov2.4sub} \\
{}^{(2)}\!{\bdelta}_{\mathrm c} &\approx - 3{}^{(2)}\!{\chi_{\mathrm c}}. \label{bdelta2_ucsub}
\end{align}
\end{subequations}
The form of these equations suggests that we define a hatted variable for
$\chi_{\mathrm c}$ according to:
\begin{equation}
{}^{(2)}\!{\hat{\chi_{\mathrm c}}} = {}^{(2)}\!{\chi_{\mathrm c}} +
3(1+c_s^2){\chi_{\mathrm c}}^2,  \label{hat_chi}
\end{equation}
in analogy with the hatted variables defined in~\eqref{hatted_var}.
On introducing these hatted variables
equations~\eqref{ucg_gov2sub} assume the following concise form:
\begin{equation} \label{ucg_gov2subX}
\partial_N {}^{(2)}\!\hat{\chi_{\mathrm c}} \approx 0,  \qquad
{\cal H}{}^{(2)}\!\hat{V}_{\mathrm c} \approx - {}^{(2)}\!\hat{\chi_{\mathrm c}},
\qquad {}^{(2)}\!\hat{\bdelta}_{\mathrm c} \approx  - 3{}^{(2)}\!\hat{\chi_{\mathrm c}}. \end{equation}
%

\section{Conserved quantities for adiabatic perturbations \label{cons}}

There are two well known conserved quantities for
long wavelength adiabatic perturbations, the curvature
perturbation in the uniform density gauge, usually denoted by $\zeta$ and the comoving
curvature perturbation, usually denoted by ${\cal R}$.
These conserved quantities were first introduced in the 1980's
for linear perturbations, $\zeta$ by Bardeen
\emph{et al} (1983)~\cite{baretal83} (see equations (2.43) and (2.45)),
and ${\cal R}$ by Bardeen (1980)~\cite{bar80}
(see equations (5.19) and (5.21)). They are defined in terms of the metric perturbations
according to\footnote{See, for example, Malik and Wands (2009)~\cite{malwan09},
equations (7.61) and (7.46), and Vernizzi (2005)~\cite{ver05}, equation (14).}
\begin{equation}
{}^{(1)}\!\zeta = -{}^{(1)}\!\psi_{\rho}, \qquad
{}^{(1)}\!{\cal R} = {}^{(1)}\!\psi_{\mathrm v}.
\end{equation}
These conserved quantities were subsequently generalized to second order.
In an important paper Malik and Wands (2004)~\cite{malwan04}
showed that ${}^{(2)}\!{\psi}_{\rho}$ is such a
conserved quantity at second order, and moreover the conservation property
depends only on the perturbed conservation of
energy equation.\footnote{See equations
(4.17), (4.18), (5.34) and (5.35) in~\cite{malwan04}.}
It is also known that the gauge invariant ${}^{(2)}\!{\psi}_{\mathrm v}$
is another conserved quantity of this type, although in this
case one has to in addition use the perturbed Einstein equations in order
to establish conservation.\footnote{See, for example, Noh and Hwang (2004)~\cite{nohhwa04},
equations (281) and (362), and Pitrou \emph{et al} (2010)~\cite{pitetal10},
equation (3.6b).}

In this section we give three new conserved quantities
at second order that are associated with the uniform curvature gauge
and relate them to the two well-known quantities. We also derive the conservation
properties in a simple, unified manner. We begin by reviewing the results at
first order, most of which are known.\footnote{An early work that considered
conserved quantities in a variety of gauges is Hwang (1994)~\cite{hwa94b} (see equations (92) and (93)).
In addition to the gauges in this paper he also uses the uniform expansion gauge, but he does not
include the gauge invariants $\chi_{\mathrm c}$ and ${\cal H}V_{\mathrm c}.$}
At first order the five gauge invariants $\psi_{\rho}$,
$\psi_{\mathrm v}$, ${\chi_{\mathrm c}}$,
$-{\cal H}{V}_{\mathrm c}$, $-\sfrac13\bdelta_{\mathrm c}$,
 are conserved for
adiabatic perturbations on super-horizon scales, and all are equal on super-horizon scales,
\begin{equation}
{}^{(1)}\!\psi_{\rho} \approx {}^{(1)}\!\psi_{\mathrm v} \approx {}^{(1)}\!{\chi_{\mathrm c}}
\approx -{\cal H}{}^{(1)}\!{V}_{\mathrm c} \approx
 -\sfrac13{}^{(1)}\!{\bdelta}_{\mathrm c}, \label{cons_summary1}
\end{equation}
the common value being the spatial function ${}^{(1)}\!C$ in the solutions in
section~\ref{solutions} below.\footnote{Some pairs are in
fact equal on all scales as indicated by $=$ rather
than $\approx$, as follows
$\psi_{\mathrm v} = {\chi_{\mathrm c}} =
-{\cal H}{V}_{\mathrm c}$,
$\psi_{\rho} =  -\sfrac13{\bdelta}_{\mathrm c}
\approx - {\cal H}{V}_{\mathrm c}$.}

At second order we have an analogous result provided one uses the gauge invariants
that correspond to the hatted variables defined in equations~\eqref{hatted_var}.
Specifically, \emph{the following gauge invariants are conserved and have the same value
for adiabatic perturbations on super-horizon scales}:
\begin{equation}
{}^{(2)}\!{\hat\psi}_{\rho} \approx {}^{(2)}\!{\hat\psi}_{\mathrm v} \approx
{}^{(2)}\!{\hat{\chi}_{\mathrm c}} \approx -{\cal H}{}^{(2)}\!{\hat{V}}_{\mathrm c} \approx
-\sfrac13{}^{(2)}\!{\hat{\bdelta}}_{\mathrm c} ,
\label{cons_summary2}
\end{equation}
the common value being the spatial function ${}^{(2)}\!C$ in the solutions in
section~\ref{solutions} below. This statement is one of the main results of this paper.

We now give a derivation of the conservation property of these quantities,
and establish the relations between them.
First, we need the perturbed energy conservation equation
in the super-horizon regime, equations~\eqref{cons_energy.1}
and~\eqref{cons_energy.2.hat}, which we specialize
to adiabatic perturbations (${}^{(r)}\!{\Gamma} \approx 0$, $r=1,2$):
\begin{equation}  \label{cons_energy_SH}
\partial_N({}^{(1)}\!{\bdelta} - 3{}^{(1)}\!{\psi}) \approx 0, \qquad
\partial_N({}^{(2)}\!\hat{\bdelta} - 3{}^{(2)}\!\hat{\psi}) \approx 0.
\end{equation}
Second, in the uniform curvature gauge two of the perturbed Einstein
equations  are constraint equations for ${\cal H}{}^{(2)}\!{V}_{\mathrm c}$
and ${}^{(2)}\!{\bdelta}_{\mathrm c}$ given in the
super-horizon regime for adiabatic perturbations
by equations~\eqref{ucg_gov2subX}, which we repeat here:
\begin{equation} \label{ucg_gov2subX.new}
{\cal H}{}^{(2)}\!\hat{V}_{\mathrm c} \approx - {}^{(2)}\!\hat{\chi_{\mathrm c}}, \qquad
{}^{(2)}\!\hat{\bdelta}_{\mathrm c} \approx  - 3{}^{(2)}\!\hat{\chi_{\mathrm c}}.
\end{equation}
Third, we specialize the constraint equation~\eqref{delta2_gen_SH} in
Appendix \ref{app:density} for ${}^{(r)}\!{\bdelta}$, $r=1,2$,
in the super-horizon regime, to the total matter gauge (${}^{(r)}\!V=0$, $r=1,2$),
which leads to
\begin{equation} \label{delta_v_zero}
{}^{(1)}\!{\bdelta}_{\mathrm v} \approx 0, \qquad
{}^{(2)}\!{\bdelta}_{\mathrm v} \approx 0.
\end{equation}
In other words, in the super-horizon regime the density
perturbations to second order in the total matter gauge are negligible
(irrespective of whether the perturbations are adiabatic).

We begin by specializing equations~\eqref{cons_energy_SH} successively to the uniform
density gauge, $\bdelta=0$, the uniform curvature gauge, $\psi=0$,
the total matter gauge, $V=0$, and conclude that ${}^{(2)}\!{\hat\psi}_{\rho}$,
${}^{(2)}\!{\hat{\bdelta}}_{\mathrm c}$ and ${}^{(2)}\!{\hat\psi}_{\mathrm v}$
are conserved, where the last result also requires the property~\eqref{delta_v_zero}.
It now follows from~\eqref{ucg_gov2subX} that
${}^{(2)}\!{\hat{\chi}_{\mathrm c}}$ and
${\cal H}{}^{(2)}\!{\hat{V}}_{\mathrm c}$ are also conserved.
We note that conservation of the
gauge invariants $\psi_{\rho}$ and
$\bdelta_{\mathrm c}$ depends only on conservation of energy
while conservation of the other gauge invariants
in~\eqref{cons_summary2} also requires the  Einstein
equations.
Continuing, the previous manipulations also establish the approximate equality
of  ${}^{(2)}\!\hat{\chi_{\mathrm c}}, -{\cal H}{}^{(2)}\!\hat{V}_{\mathrm c}$
and $-\sfrac13{}^{(2)}\!\hat{\bdelta}_{\mathrm c}$.
Finally we can establish that ${}^{(2)}\!\hat\psi_{\rho}$ is equal to these
variables and to ${}^{(2)}\!\hat\psi_{\mathrm v}$ by using a change
of gauge formula in the super-horizon regime, which reads:\footnote{Specialize
equation (49) in UW1~\cite{uggwai19a} to adiabatic
perturbations in the super-horizon regime, and
use $\partial_N {}^{(1)}\!\psi_{\rho}\approx 0$ to obtain
the second order formula. }
\begin{equation} \label{psi_change_rho}
{}^{(1)}\!{\psi}_{\rho} = {}^{(1)}\!{\psi} -
\sfrac13 {}^{(1)}\!{\bdelta}, \qquad
{}^{(2)}\!\hat{\psi}_{\rho} \approx {}^{(2)}\!\hat{\psi} - \sfrac13 {}^{(2)}\!\hat{\bdelta}.
\end{equation}
Choose the gauge on the right side of these
equations to be successively the uniform curvature gauge and
the total matter gauge and use~\eqref{delta_v_zero} to obtain
\begin{equation}
{}^{(2)}\!\hat{\psi}_{\rho} \approx - \sfrac13 {}^{(2)}\!\hat{\bdelta}_{\mathrm c}
\approx  {}^{(2)}\!\hat{\psi}_{\mathrm v}.
\end{equation}

It should be noted that if ${}^{(1)}\!\psi_{\rho}$ and ${}^{(2)}\!\hat{\psi}_{\rho}$
are conserved then so is the un-hatted variable ${}^{(2)}\!{\psi}_{\rho}$, because the
coefficient in the definition~\eqref{hat1} of ${}^{(2)}\!{\hat{\psi}}$ is constant. The same
remark applies to ${}^{(2)}\!{\psi}_{\mathrm v}$. However, for the other variables
in~\eqref{cons_summary2} conservation of the hatted variable does not apply
conservation of the un-hatted variable \emph{unless
$q$ and $c_s^2$ are constant.}

We end this section by pointing out that there is a special
class of perturbed FL cosmologies, namely the $\Lambda CDM$ universes,
which admit \emph{linear}
conserved quantities on super-horizon scale that remain conserved on
all scales.
Specifically, \emph{the linear comoving curvature perturbation
${}^{(1)}\!{\cal R} = {}^{(1)}\!\psi_{\mathrm v}$ is
conserved on all scales}, as are the related gauge invariants
${}^{(1)}\!\chi_{\mathrm c}=-{\cal H}{}^{(1)}\!V_{\mathrm c}={}^{(1)}\!\psi_{\mathrm v}$.
This conservation property follows from the fact that for a perturbed
$\Lambda CDM$ universe the governing equation~\eqref{totmat_1.1} reduces to
the exact equation $\partial_N {}^{(1)}\!\psi_{\mathrm v}=0$,
since $c_s^2=0$ and ${}^{(1)}\!\Gamma=0$.
On the other hand the linear  curvature perturbation in the uniform density gauge
${}^{(1)}\!{\zeta} = -{}^{(1)}\!\psi_{\rho}$ does not have this property, and neither
do any of the second order conserved quantities.

\section{The general solution for adiabatic perturbations \label{solutions}}

In this section we derive the general solution of the governing equations
for adiabatic perturbations in the super-horizon regime using the total matter gauge.
We then obtain the solution in the uniform curvature gauge and the Poisson
gauge by using the change of gauge formulas given in UW1~\cite{uggwai19a}.
The conserved quantities described in section~\ref{cons}
emerge naturally in the solution process,
beginning with ${}^{(1)}\!{ \psi}_{\mathrm v}$ and ${}^{(2)}\!{\hat \psi}_{\mathrm v}$,
and continuing with equation~\eqref{SH_uc2}.


\subsection{Solving in the total matter gauge \label{totmat_soln}}

The governing equations for linear perturbations in the
total matter gauge when specialized to adiabatic perturbations
in the super-horizon regime assume the
simple form~\eqref{totmat_sh_adiabatic}, which we repeat here
but with $N$ replaced by the  background
scale factor $a$ as time variable.
Using $\partial_N=a\partial_a$ we obtain:
\begin{subequations} \label{totmat_1}
\begin{align}
\partial_a {}^{(1)}\!{\psi}_{\mathrm v}
 &\approx 0, \label{totmat_1.2}   \\
\partial_a(a^2\,{}^{(1)}\!B_{\mathrm v} ) &\approx
a{\cal H}^{-1}{}^{(1)}\!{\psi}_{\mathrm v},  \label{totmat_1.3}
\end{align}
\end{subequations}
with
\begin{subequations} \label{totmat.soln1}
\begin{equation}
{}^{(1)}\!\phi_{\mathrm v}\approx 0,\qquad
{}^{(1)}\!\bdelta_{\mathrm v}\approx 0.    \label{comov_sh10}
\end{equation}
It follows immediately from~\eqref{totmat_1.2} that
\begin{equation}
{}^{(1)}\!\psi_{\mathrm v}\approx {}^{(1)}\!C,   \label{comov_sh1}
\end{equation}
\end{subequations}
where we identify the spatial function ${}^{(1)}\!C(x^i)$ as the conserved
quantity at first order. Solving~\eqref{totmat_1.3}
for ${}^{(1)}\!B_{\mathrm v}$ gives
\begin{equation}
{\cal H}{}^{(1)}\!B_{\mathrm v} \approx
\left(\frac{\cal H}{a^2} \int_0^a \frac{\bar a}{{\cal H}(\bar a)}d{\bar a}\right){}^{(1)}\!C
+\frac{\cal H}{a^2} {}^{(1)}\!C_{*},
\end{equation}
where ${}^{(1)}\!C_{*}=\lim_{a\rightarrow 0} a^2 \,{}^{(1)}\!B_{\mathrm v}$ is a second
arbitrary spatial function.
In terms of the perturbation growth function
$g(a)$ defined in equation~\eqref{def_g_simple} we obtain
\begin{equation}
{\cal H}{}^{(1)}\!B_{\mathrm v} \approx (1-g){}^{(1)}\!C +\frac{\cal H}{a^2} {}^{(1)}\!C_{*},
 \label{comov_sh1_B}
\end{equation}
which with~\eqref{totmat.soln1} gives the general solution
at first order.

We make a brief remark  on the physical viability of the solution.
We assume that the deceleration parameter satisfies
the weak restriction $q>-2$, which
implies that ${\cal H}/a^2$ is a decreasing function
and that ${\cal H}/a^2\rightarrow \infty$ as $a\rightarrow 0.$ We thus refer to term
$({\cal H}/a^2) {}^{(1)}\!C_{*}$ in the
solution as the decaying mode.  If the decaying mode is present
 (${}^{(1)}\!C_{*}\neq 0$) we impose a restriction of the form $a>a_{*}>0$
 on the time evolution in order to ensure that the decaying mode is sufficiently
 small in the time period under consideration. Since the
 decaying mode enters into ${}^{(1)}\!\bdelta_{\mathrm v}$ this restriction
 is necessary to
 ensure that ${}^{(1)}\!\bdelta_{\mathrm v}\approx 0$
 is valid in some time interval $a>a_{*}$.\footnote
 {Martin and Schwarz (1998)~\cite {marsch98} do not impose a restriction
 of the form $a>a_{*}$ and hence argue that the decaying mode has to be excluded
 (see the remark following their equation (4.10)).}

At second order the governing equations for adiabatic
perturbations on super-horizon scales in
the total matter gauge are given by equations~\eqref{totmat_sh_adiabatic},
which we repeat here:
\begin{subequations}\label{totmat_2.repeat}
\begin{align}
\partial_a {}^{(2)}\!\hat{\psi}_{\mathrm v} &\approx0,  \label{totmat_2.2.repeat}  \\
\partial_a(a^2\,{}^{(2)}\!B_{\mathrm v} ) &\approx
a{\cal H}^{-1}({}^{(2)}\!\hat{\psi}_{\mathrm v}
- 2{\mathbb D}_0({}^{(1)}\!\psi_{\mathrm v}) ),   \label{totmat_2.3.repeat}
\end{align}
\end{subequations}
with
\begin{subequations} \label{comov_sh2}
\begin{equation}
{}^{(2)}\!\phi_{\mathrm v}\approx 0, \qquad
{}^{(2)}\!{\bdelta}_{\mathrm v}\approx 0,  \label{comov_sh2.1}
\end{equation}
where ${\mathbb D}_0$ is defined in~\eqref{D_0}.
We write the solution of~\eqref{totmat_2.2.repeat} as
\begin{equation}
{}^{(2)}\!\hat{\psi}_{\mathrm v}\approx {}^{(2)}\!C, \label{psi_v2}
\end{equation}
where we identify the spatial function ${}^{(2)}\!C(x^i)$ as the conserved
quantity at second order.
Observe that \emph{the differential equation~\eqref{totmat_2.3.repeat}
for ${}^{(2)}\!B_{\mathrm v}$ is essentially
the same as equation~\eqref{totmat_1.3} for ${}^{(1)}\!B_{\mathrm v}$}, with the
spatial function ${}^{(1)}\!C$  on the right side replaced by the spatial function
${}^{(2)}\!C - 2{\mathbb D}_0({}^{(1)}\!C)$.
It follows immediately on taking note of equation~\eqref{comov_sh1_B}
that the solution for ${}^{(2)}\!B_{\mathrm v}$ is
\begin{equation} \label{B_v2}
{\cal H}{}^{(2)}\!B_{\mathrm v} \approx\,(1-g)\left({}^{(2)}\!C -
2{\mathbb D}_0( {}^{(1)}\!C)\right) +\frac{\cal H}{a^2} {}^{(2)}\!C_{*},
\end{equation}
\end{subequations}
where ${}^{(2)}\!C_{*}$ represents the decaying mode at second order.
\emph{Equations~\eqref{comov_sh2} give the general solution
at second order, including the decaying mode, in the total matter gauge}.
If ${}^{(2)}\!C_{*}\neq 0$ a restriction of the form $a>a_{*}>0$ is
again needed.

\subsection{Transforming to the uniform curvature gauge \label{uc_soln}}

The link with the uniform curvature gauge at
first order is provided by the following change of
gauge formulas UW1~\cite{uggwai19a}:
\begin{subequations} \label{link_uc}
\begin{equation}
{\cal H}V_{\mathrm c}=-\psi_{\mathrm v}, \qquad
{\cal H}B_{\mathrm c} ={\cal H}B_{\mathrm v} -\psi_{\mathrm v},
\end{equation}
where we are dropping the superscript
${}^{(1)}$ on the linear solution.
We also need the density and velocity constraints~\eqref{ucg_gov1sup}
at first order which read
\begin{equation}
{\cal H}V_{\mathrm c}= -\chi_{\mathrm c}, \qquad
{\bdelta}_{\mathrm c}\approx 3 {\cal H}V_{\mathrm c}.
\end{equation}
\end{subequations}
It follows from~\eqref{comov_sh1} and~\eqref{comov_sh1_B} using~\eqref{link_uc} that
\begin{subequations}  \label{uc_soln1}
\begin{equation}
\chi_{\mathrm c}\approx C, \qquad {\cal H}V_{\mathrm c}\approx - C,
\qquad  {\bdelta}_{\mathrm c}\approx -3 C, \qquad
{\cal H}B_{\mathrm c}\approx -gC +\frac{\cal H}{a^2}C_{*},   \label{SH_uc1}
\end{equation}
while by~\eqref{def_chi} we obtain
\begin{equation}
\phi_{\mathrm c}=(1+q)\chi_{\mathrm c} \approx (1+q)C,
\end{equation}
\end{subequations}
which give the linear perturbations in the uniform curvature gauge.

The link with the uniform curvature gauge at
second order is provided by the following change of
gauge formulas:
\begin{subequations}
\begin{align}
{\cal H}{}^{(2)}\!{\hat V}_{\mathrm c}&\approx  -
{}^{(2)}\!{\hat\psi}_{\mathrm v},  \label{V_c_change}  \\
{\cal H}{}^{(2)}\!\hat{B}_\mathrm{c} &\approx {\cal H}{}^{(2)}\!\hat{B}_{\mathrm v} -
{}^{(2)}\!{\hat\psi}_{\mathrm v} +
2\partial_N ({\cal H}B_\mathrm{v})\psi_{\mathrm v}  -
{\cal H}B_{rem,\mathrm{v},\mathrm{c}}, \label{B_c_change}
\end{align}
\end{subequations}
given by equations~\eqref{change_uc1}
and~\eqref{gc_Bv2} in Appendix~\ref{app:changegauge}.
We also need the density and velocity constraints~\eqref{ucg_gov2subX}
which read
\begin{equation} \label{ucg2_constraints}
{\cal H}{}^{(2)}\!\hat{V}_{\mathrm c} \approx - {}^{(2)}\!\hat{\chi_{\mathrm c}} , \qquad
{}^{(2)}\!\hat{\bdelta}_{\mathrm c} \approx  - 3{}^{(2)}\!\hat{\chi_{\mathrm c}}.
\end{equation}
%
%
It immediately follows from~\eqref{psi_v2}, \eqref{V_c_change}
and~\eqref{ucg2_constraints} that
\begin{subequations}\label{uc_sh2}
\begin{equation}
{}^{(2)}\!\hat{\chi_{\mathrm c}}\approx {}^{(2)}\!C, \qquad
{\cal H}{}^{(2)}\!\hat{V}_{\mathrm c} \approx - {}^{(2)}\!C, \qquad
{}^{(2)}\!\hat{\bdelta}_{\mathrm c} \approx  - 3{}^{(2)}\!C,  \label{SH_uc2}
\end{equation}
where the spatial function ${}^{(2)}\!C(x^i)$ is the conserved
quantity at second order. The metric perturbation ${}^{(2)}\!\phi_{\mathrm c}$
is determined by first finding ${}^{(2)}\!{\chi_{\mathrm c}}$
using~\eqref{hat_chi} and then using the definition~\eqref{def_chi},
which leads to
\begin{equation} \label{phi_c.2}
{}^{(2)}\!\phi_{\mathrm c} \approx (1+q) \left({}^{(2)}\!C +
\left( 2(1+q) +3( w-c_s^2)\right )C^2\right).
\end{equation}
Note that the decaying mode does not enter into the expressions~\eqref{SH_uc2}
and~\eqref{phi_c.2}.

We finally use~\eqref{B_c_change} in conjunction
with~\eqref{psi_v2} and~\eqref{B_v2} and the definitions of the hatted
variables~\eqref{hatted_var} to obtain an expression
for ${\cal H}{}^{(2)}\!B_{\mathrm c} $. This necessitates using the
first order solution that is given
by~\eqref{totmat.soln1},~\eqref{comov_sh1_B},~\eqref{uc_soln1}
and~\eqref{link_poisson} to
evaluate the complicated  source term ${\cal H}B_{rem,\mathrm{v},\mathrm{c}}$
given by equation~\eqref{rem_B} in Appendix~\ref{app:changegauge}.
 At this stage, in the interests
of simplicity, we drop the decaying mode.
The final result is
\begin{equation}\label{SH_uc3}
{\cal H}{}^{(2)}\!B_{\mathrm c} \approx -g\,{}^{(2)}\!C +
\left(g -(1+q)(g+1) \right)C^2 +
2(q-1)g{\mathbb D}_0(C).
\end{equation}
\end{subequations}
In summary equations~\eqref{uc_sh2} give the solution at second
order in the uniform curvature gauge, with the decaying mode set to zero
in~\eqref{SH_uc3}. If needed the decaying mode terms can be worked out
without difficulty.

\subsection{Transforming to the Poisson gauge \label{Poisson_soln}}

It turns out that the super-horizon solution has its most complicated form
when expressed in the Poisson gauge. At first order the link with the Poisson gauge
is provided by the following change of gauge
formulas (UW1~\cite{uggwai19a}, section 3):
\begin{subequations}\label{link_poisson}
\begin{equation}
\psi_{\mathrm p}=\psi_{\mathrm v} -{\cal H}B_{\mathrm v}, \qquad
{\cal H}V_{\mathrm p}=-{\cal H}B_{\mathrm v}, \qquad
{\bdelta}_{\mathrm p}= {\bdelta}_{\mathrm v} -3{\cal H}B_{\mathrm v},
\label{poisson-comov1}
\end{equation}
and the perturbed Einstein equations give
\begin{equation}
\phi_{\mathrm p}=\psi_{\mathrm p}.
\end{equation}
\end{subequations}
It follows from~\eqref{totmat.soln1} and~\eqref{comov_sh1_B}
using~\eqref{link_poisson} that
\begin{equation}
\psi_{\mathrm p} \approx g C -\frac{\cal H}{a^2} C_{*},
\quad {\cal H}{V}_{\mathrm p} \approx -(1-g)C -\frac{\cal H}{a^2} C_{*}, \quad
{\bdelta}_{\mathrm p} \approx -3(1-g)C -3\frac{\cal H}{a^2} C_{*},  \label{SH_poisson1}
\end{equation}
which give the linear perturbations in the Poisson gauge.

The link with the Poisson gauge at second order is obtained by
generalizing the change of gauge
formulas~\eqref{poisson-comov1} to second order, as in
equations~\eqref{poisson1},~\eqref{poisson2} and~\eqref{poisson3}.
We use~\eqref{poisson1} to first calculate
${}^{(2)}\!{\hat \psi}_{\mathrm p}$ in terms of ${}^{(2)}\!{\hat \psi}_{\mathrm v}$
and ${}^{(2)}\!{\hat B}_{\mathrm v}$, and then set $\phi_{\mathrm v}=0$
in~\eqref{poisson2} and~\eqref{poisson3}
to get ${\cal H}{}^{(2)}\!{\hat V}_{\mathrm p}
\approx {}^{(2)}\!{\hat\psi}_{\mathrm p} - {}^{(2)}\!C$ and
${}^{(2)}\!{\hat{\bdelta}}_{\mathrm p} \approx \,3{\cal H}{}^{(2)}\!{\hat V}_{\mathrm p}$.
The only use of the
perturbed Einstein equations is to relate ${}^{(2)}\!{\hat \phi}_{\mathrm p}$
to ${}^{(2)}\!{\hat \psi}_{\mathrm p}$ as in equation~\eqref{poisson4}.
The results for the unhatted variables, obtained
using~\eqref{hatted_var}, are as follows:
\begin{subequations} \label{poisson_sh2}
\begin{align}
{}^{(2)}\! \psi_{\mathrm p} &\approx \,g{}^{(2)}\!C +
\left((1+q)(1-g)^2 - g^2 - g\right) C^2 +
4g(1-g){\mathbb D}_0(C), \label{poisson_2.1} \\
{}^{(2)}\! \phi_{\mathrm p} &\approx \,{}^{(2)}\! \psi_{\mathrm p} +4g^2C^2 -
4\left((1+q)(1-g)^2 +g^2\right) {\mathbb D}_0(C),\label{poisson_2.2}  \\
{\cal H}{}^{(2)}\!{ V}_{\mathrm p} &\approx
-(1-g){}^{(2)}\!C -g(1-g)\left(C^2 -4{\mathbb D}_0(C)\right),
\label{poisson_2.3}\\
{}^{(2)}\!{\bdelta}_{\mathrm p} &\approx \,3{\cal H}{}^{(2)}\! V_{\mathrm p}
+9[(1+c_s^2)+\sfrac12(1+w)]C^2,
\label{poisson_2.4}
\end{align}
\end{subequations}
with the decaying mode set to zero
(${}^{(r)}\!C_{*}=0, \,r=1,2$) in the interest of simplicity. Note
that the decaying mode would appear in each of these expressions.

\section{Applications \label{applications}}

The solution of the governing equations for adiabatic long-wavelength perturbations given
in section~\ref{totmat_soln}  using the total
matter gauge (see equations~\eqref{comov_sh2}) is general in the sense that it is valid
for any stress-energy tensor of the form~\eqref{Tab} (zero
anisotropic stress and heat flux), and also includes the
decaying mode. The spatial dependence of
the solution is determined by four spatial functions,
the two functions ${}^{(1)}\!C$ and ${}^{(2)}\!C$, which determine
the growing mode and represent the
conserved quantities, and the two functions ${}^{(1)}\!C_{*}$
and ${}^{(2)}\!C_{*}$, which determine the decaying mode.
The dependence in time of the growing mode at first
and second order is
determined solely by the perturbation growth function $g(a)$. Indeed the solution
as derived in the total matter
gauge has a remarkably simple form. In the uniform curvature gauge
(see equations~\eqref{uc_sh2}) and  Poisson gauge
(see equations~\eqref{poisson_sh2}), however, the perturbations
at second order also depend on the matter variables $w$ and $c_s^2$.

Before giving some examples we briefly digress to relate
the arbitrary functions ${}^{(r)}\!C$, $r=1,2$, to
the usual conserved quantities ${}^{(r)}\!\zeta\equiv -{}^{(r)}\!\psi_{\rho}$
and ${}^{(r)}\!{\cal R} \equiv {}^{(r)}\!\psi_{\mathrm v}$, which
are approximately equal but opposite in sign for adiabatic perturbations in the super horizon regime. In our
derivation of the solutions we introduced
${}^{(1)}\!C$ as ${}^{(1)}\!\psi_{\mathrm v} $, and ${}^{(2)}\!C$ as
${}^{(2)}\!{\hat\psi}_{\mathrm v} $. It follows that
\begin{subequations}
\begin{align}
{}^{(1)}\!C&\equiv {}^{(1)}\!{\cal R}\approx - {}^{(1)}\!\zeta, \\
{}^{(2)}\!C&\equiv {}^{(2)}\!{\cal R} +2{}^{(1)}\!{\cal R}^2\approx
 -( {}^{(2)}\!\zeta -2{}^{(1)}\!\zeta^2),  \label{C_link}
\end{align}
\end{subequations}
since ${}^{(2)}\!{\hat\psi }={}^{(2)}\!\psi + 2{}^{(1)}\!\psi^2.$
We mention that in inflationary cosmology it is customary to
parametrize the primordial non-Gaussianity level in terms of
the conserved curvature perturbation $\zeta$ according to
\begin{equation}
{}^{(2)}\!\zeta=2a_{NL} {}^{(1)}\!\zeta^2, \label{aNL}
\end{equation}
where the parameter $a_{NL}$ depends on the physics
of the type of inflation (see, for example, Bartolo \emph{et al}
(2010)~\cite{baretal10}, equation (38)). In terms of our
conserved quantity $C$ the relation~\eqref{aNL} reads
\begin{equation}
{}^{(2)}\!C=2(1-a_{NL}) {}^{(1)}\!C^2.
\end{equation}
For standard single field inflation $a_{NL}\approx 1$ and hence ${}^{(2)}\!C=0$ .

The general solution that we derived in section~\ref{totmat_soln} applies to the case
of a perfect fluid with a barotropic
equation of state $p = p(\rho)$ since then the adiabaticity conditions
${}^{(r)}\!\Gamma = 0$, $r=1,2$, are satisfied. In this case the scalars $w$ and $c_s^2$
are determined by the equation of state.
In the special case of a linear equation of state $p=w\rho$ with $w$ constant
and $w>-\frac53$, it follows that $q>-2$ is constant and integrating
$a\partial_a{\cal H} = -q{\cal H}$ gives
\begin{equation}
{\cal H}(a)={\cal H}_0 (a/a_0)^{-q},
\end{equation}
where ${\cal H}_0={\cal H}(a_0),$ where $a_0$ is a fixed reference epoch.\footnote{Here
and in the rest of this section we are temporarily
suspending our convention of using ${}_0$ to denote
a background quantity and are instead using it to refer to the value of some quantity
at a fixed reference epoch denoted by $a_0$.}
 On
substituting this expression in the definition~\eqref{def_g_simple} of the
perturbation growth function $g(a)$ we obtain
\begin{equation}
g(a) =\frac{1+q}{2+q} = \frac{3(1+w)}{5+3w},
\end{equation}
\emph{i.e.} $g(a)$ is constant.
Note that $g(a)=\sfrac35$ for dust and $g(a)=\sfrac23$ for radiation.
In this case the solution in the Poisson gauge given
by~\eqref{poisson_2.1} and~\eqref{poisson_2.2} simplifies considerably, resulting in
\begin{subequations} \label{poisson_2}
\begin{align}
{}^{(2)}\!{\hat \psi}_{\mathrm p} &\approx \,g{}^{(2)}\!C +
4g(1-g){\mathbb D}_0( {}^{(1)}\!C), \label{poisson_2.1a} \\
{}^{(2)}\!{\hat \phi}_{\mathrm p} &\approx \,g{}^{(2)}\!C -
4g^2{\mathbb D}_0({}^{(1)}\!C). \label{poisson_2.2b}
\end{align}
\end{subequations}

The general solution also applies to long wavelength perturbations in
a two-fluid universe with the matter described as a
single fluid with barotropic equation of state, so that the
perturbations are adiabatic. The two fluids are assumed to be non-interacting
each with a linear equation of state, with parameters $w_1,w_2$ satisfying $w_2<w_1$.
Two cases of particular interest are the radiation-matter universe
with $w_2=0,w_1=\sfrac13$ and the $\Lambda CDM$ universe with
$w_2=-1, w_1=0$. The former case arises when deriving an expression
for the second order early Integrated Sachs-Wolfe effect in the anisotropy of the CMB
on large scales (Bartolo \emph{et al}  (2006)~\cite{baretal06},  equations (3.9)-(3.10),
Section IIIC, and Appendix C.)

In order to calculate $g(a)$ we need an expression
for ${\cal H}(a)$. Conservation of energy for each fluid leads to
$\rho_A/\rho_{A,0}=x^{-3(1+w_A)}, x=a/a_0, A=1,2$, where
$\rho_A, A=1,2$ are the background densities of the fluids
and $\rho_{A,0}=\rho_A(a_0).$
It follows that the individual density
parameters $\Omega_A=\rho_A/(3H^2), A=1,2$ are given by
\begin{equation} \label{omega_A}
\Omega_A=\Omega_{A,0}x^{-(1+3w_A)}\left(\frac{{\cal H}_0}{\cal H}\right)^2,
\end{equation}
where $\Omega_{A,0}=\rho_{A,0}/(3H_0)^2, A=1,2$.
Since the background is flat, we have $\Omega_1+\Omega_2=1$
and~\eqref{omega_A} leads to
\begin{equation} \label{2fluidH}
\left(\frac{\cal H}{{\cal H}_0}\right)^2=\Omega_{1,0}x^{-(1+3w_1)} +
\Omega_{2,0}x^{-(1+3w_2)}, \quad x=a/a_0,
\end{equation}
where $\Omega_{1,0}+\Omega_{2,0}=1$.
We can now substitute~\eqref{2fluidH} in~\eqref{def_g_simple} to
obtain an explicit expression for $g(a)$ which determines all the
first order perturbations, and in the case of the total matter gauge,
also the second order perturbations. The matter parameters
 $w$ and $c_s^2$ for the combined fluid are given by:
\begin{equation} \label{2fluidw}
w=w_1\Omega_1 +w_2\Omega_2,\qquad
c_s^2=\frac{w_1(1+w_1)\Omega_1+w_2(1+w_2)\Omega_2}{1+w},
\end{equation}
where the $\Omega_A$ are given by~\eqref{omega_A}. As an example the
curvature perturbation $\psi_{\mathrm p}$ in the Poisson gauge is given by
equations~\eqref{poisson_2.1} and~\eqref{hat1}:
\begin{subequations} \label{psi_p.gen}
\begin{equation}
{}^{(1)}\!{ \psi}_{\mathrm p} \approx \,g{}^{(1)}\!C,
\end{equation}
\begin{equation} \label{psi_p.gen2}
{}^{(2)}\!{ \psi}_{\mathrm p} \approx \,g{}^{(2)}\!C +
\left(\sfrac32(1+w)(1-g)^2 - g^2 - g \right) {}^{(1)}\!C^2 +
4g(1-g){\mathbb D}_0( {}^{(1)}\!C).
\end{equation}
\end{subequations}
At second order the leading order term is determined by $g$ alone while
the source terms depend also on $w$.

For all values of $w_1$ and $w_2$ it has been shown
by Hu and Eisenstein (1999)~\cite{hueis99} that the integral
in~\eqref{def_g_simple} that determines $g$ for
these two-fluid models can be expressed in terms
of the incomplete beta function, and that if $(5+3w_1)/3(w_1-w_2)$ is an integer
then $g(a)$ can be expressed in elementary form (see page 12 in~\cite{hueis99}).
We now consider a radiation-matter universe ($w_1=\sfrac13, w_2=0$),
which satisfies this condition.

 In this case it is convenient to choose $a_0=a_{eq}$, the epoch of
 matter-radiation equality. It follows that $\Omega_{1,0}=\Omega_{2,0}=\sfrac12$,
and~\eqref{2fluidH} simplifies to give
\begin{equation}
{\cal H}(a)={\cal H}_{eq}\frac{\sqrt{ x+1}}{\sqrt {2}x}, \qquad  x=a/a_{eq},
\end{equation}
It is a simple matter to evaluate the integral~\eqref{def_g_simple}
for $g(a)$ to obtain
\begin{subequations} \label{g_matrad}
\begin{equation} \label{g_radmat}
g(a)=\sfrac{1}{15}x^{-3} ( 9x^3+2x^2-8x-16 +16\sqrt{1+x}), \qquad  x=a/a_{eq}.
\end{equation}
In addition~\eqref{2fluidH} and~\eqref{omega_A} lead directly to
\begin{equation}
w=\frac{1}{3(1+x)} \qquad 3c_s^2 = \frac{4}{3x+4}.
\end{equation}
\end{subequations}
As expected it follows that $\lim_{a\rightarrow 0}g(a)=\sfrac23$
(radiation) and
$\lim_{a\rightarrow \infty}g(a)=\sfrac35$ (pressure-free matter).
The curvature perturbation $\psi_{\mathrm p}$ in the Poisson
gauge, given by~\eqref{psi_p.gen}, can now be calculated
using~\eqref{g_matrad}.
The first order expression has been given, for
example, by Hu and Eisenstein (1999)~\cite{hueis99}
(see equation (67)).\footnote
{This expression for $g(a)$ has also been given by
Kodama and Sasaki (1984)~\cite{kodsas84} (see equations
(IV.4.11) and (IV.4.14) with $z=1+x$), Dodelson (2003)~\cite{dod03}
(see equation (7.32), up to a constant multiplicative factor),
and Bartolo \emph{et al} (2006)~\cite{baretal06} (see equation (5.19)).
Mukhanov (2005)~\cite{muk05} gives
an expression for $g(\eta)$, see equation (7.71). }
To the best of our knowledge the second order expression
is new.\footnote
{An expression for ${}^{(2)}\!{ \psi}_{\mathrm p}$ for
the radiation-matter universe has been given by
Bartolo \emph{et al}  (2006)~\cite{baretal06} (see equation (3.48)),
but the source term was left as a complicated integral.}

The second special case of importance is the perturbed $\Lambda CDM$
universe given by $w_2=-1, w_1=0$. It follows from~\eqref{2fluidH}
that
\begin{equation}  \label{H_lambdaCDM}
{\cal H}^2 =  {\cal H}_0^2\left(\Omega_{m,0}x^{-1} + \Omega_{\Lambda,0}x^{2}\right)
\qquad x=a/a_0,
\end{equation}
which when substituted into~\eqref{def_g_simple} gives $g(a)$
for the $\Lambda CDM$ universe.\footnote
{We note that the function  $g(a)$ for $\Lambda CDM$ can
be represented in different ways and has been
studied extensively, as described in section~\ref{function_g}
(see equations~\eqref{D_LCDM} and~\eqref{g_D}).}
From~\eqref{omega_A} and~\eqref{2fluidw} we obtain
\begin{equation}  \label{1+w,LCDM}
1+w=\Omega_m=\Omega_{m,0}x^{-1} \left(\frac{{\cal H}_0}{\cal H}\right)^2.
\end{equation}
With these expressions one can use~\eqref{psi_p.gen} to calculate
the long wavelength curvature perturbation $\psi_{\mathrm p}$
in the Poisson gauge, and any other
perturbations for the $\Lambda CDM$ universe
using the results of section~\ref{solutions}.
In this case, however, one can do more: since $c_s^2=0$ and $\Gamma=0$
for the perturbed $\Lambda CDM$ universe
the full (\emph{i.e.} non-truncated)
equations~\eqref{totmat_1_new} at linear order in the total matter
gauge can be solved explicitly as in the super-horizon case, giving
the \emph{exact} expressions
\begin{equation} \label{totmat,exact}
\psi_{\mathrm v} = {}^{(1)}\!C, \qquad \phi_{\mathrm v} =0, \qquad
{\cal H}{B}_{\mathrm v}=(1-g){}^{(1)}\!C,
\end{equation}
where ${}^{(1)}\!C$ is the conserved quantity.
The new feature is the exact expression for the density perturbation which
 we can calculate using
\begin{equation}
{\bdelta}_{\mathrm v} =
\sfrac23(1+w)^{-1}{\cal H}^{-2}{\bf D}^2 ({\psi}_{\mathrm v} -
{\cal H}{B}_{\mathrm v}).
\end{equation}
It follows from~\eqref{1+w,LCDM} and~\eqref{totmat,exact} that
\begin{equation}
{\bdelta}_{\mathrm v}= \sfrac23 m^{-2} xg{\bf D}^2 {}^{(1)}\!C, \qquad x=a/a_0,
\end{equation}
 where $m^2$ is a constant given
 by $m^2={\cal H}_0^2\,\Omega_{m,0}.$

Furthermore, the full (non-truncated)
equations~\eqref{totmat_2_new} at second order in the total matter
gauge can likewise be solved explicitly, and one finds that
the evolution of the perturbations
${}^{(2)}\!\psi_{\mathrm v},  {}^{(2)}\!\phi_{\mathrm v},
{\cal H}{}^{(2)}\!{B}_{\mathrm v}$ and
${}^{(2)}\!{\bdelta}_{\mathrm v}$
 is again determined by $g(a)$,
partly algebraically and partly through an integral involving $g(a)$.
We will give details elsewhere.
We note, however, that the density perturbation
${}^{(2)}\!{\bdelta}_{\mathrm v}$ has been
previously determined in an indirect way  and this expression shows the role
played by $g(a)$ (see Uggla and Wainwright (2014)~\cite{uggwai14b},
equations (10), (13) and (16)). Our simple method of integration
using the total matter gauge confirms the earlier result.

\section{The perturbation evolution function $g$ \label{function_g} }

The function $g(a)$ is defined by equation~\eqref{def_g_simple}, which we repeat
here:
\begin{equation}
g(a) = 1 - \frac{\cal H}{a^2} \int_0^a \frac{\bar a}{{\cal H}(\bar a)}d{\bar a}.  \label{def_g_simple_repeat}
\end{equation}
This function first emerged in this paper when we solved the governing equations in
the total matter gauge at first order to obtain the metric perturbation $B_{\mathrm v}$.
We subsequently showed  that it determines
the evolution of the perturbations at first order in all the standard
gauges. In particular, in the Poisson gauge which plays an important role in applications,
$g$ determines the growing mode of the curvature
perturbation $\psi_{\mathrm p}$ at first order
in the long wavelength limit according to\footnote{This follows
from~\eqref{comov_sh1} and~\eqref{SH_poisson1},
noting that $\psi_{\mathrm v}={\cal R}.$   }
\begin{equation} \label{psi/R}
\psi_{\mathrm p}/{\cal R}\approx g.
\end{equation}
In other words $g$ represents the
growth of the non-conserved Poisson curvature perturbation $\psi_{\mathrm p}$
relative to the conserved comoving curvature perturbation ${\cal R}$.
We note that the ratio $\psi_{\mathrm p}/{\cal R}$ has been emphasized by
Hu and Eisenstein (1999)~\cite{hueis99}, who
derived the following expression for long wavelength adiabatic perturbations
with negligible anisotropic stress:\footnote
{See equation (59) in~\cite{hueis99}, dropping the decaying mode,
neglecting the second term
and noting that $\Phi$ and $\zeta$  correspond to
our $\psi_{\mathrm p}$ and ${\cal R}$.}
\begin{equation} \label{psi/R,hueis}
\psi_{\mathrm p}/{\cal R}\approx
1-\frac{\sqrt{\rho}}{a}\int_0^a\frac{d\bar {a}}{\sqrt{\rho({\bar a})}},
\end{equation}
where $\rho$ denotes the background matter density.
The relation $\rho=3H^2$, valid in a flat background, shows that
the integral in~\eqref{psi/R,hueis} is equal to the integral
in~\eqref{def_g_simple_repeat}.

We now derive some properties of $g$, first noting that $g$
can also be expressed as a function of $t$ or of $\eta$
by making a change of variable in the integral, leading to:
\begin{equation} \label{g_alt}
g(t)=1-\frac{H}{a}\int_0^t a(\bar t) d{\bar t}, \qquad
g(\eta)=1- \frac{\cal H}{a^2}\int _0^{\eta}a(\bar \eta)^2 d{\bar \eta}.
\end{equation}
The initial singularity is given by $a=0$, with the
clock time  translated so that $t=0$ when $a=0$.
We assume that $H>0$ and that $q>-2$ for all $t>0$.
It follows from the first of equations~\eqref{g_alt}
that $g(t)<1$  for $t>0$.

As regards asymptotic behaviour, if
$H/a\rightarrow \infty$, $q\rightarrow q_{sing}$  as $t\rightarrow 0$ and
$H/a\rightarrow 0,$ $q\rightarrow q_\infty$  as $t\rightarrow \infty$,
where $q_{sing},  q_\infty >-2$, then it follows
from the first of equations~\eqref{g_alt} that\footnote{Write
$g(t)=1-\frac{\int_0^t a(\bar t) d{\bar t}}{a/H}$ and apply
l'H\^{o}pital's rule to the indeterminate ratio using~\eqref{deriv.a/H}.}
\begin{equation} \label{lim.g(t)}
\lim_{t\rightarrow 0}g(t) = \frac{1+q_{sing}}{2+q_{sing}}=
\frac{3(1+w_{sing})}{5+3w_{sing}}, \qquad
\lim_{t\rightarrow \infty}g(t) = \frac{1+q_\infty}{2+q_\infty}=\frac{3(1+w_\infty)}{5+3w_\infty},
\end{equation}
are finite.
By integrating the identity
\begin{equation} \label{deriv.a/H}
\partial_t\left(\frac{a}{H}\right)-a=a(1+q),
\end{equation}
we can write $g(t)$  in the alternate form
\begin{equation} \label{def_g_alt}
g(t)= \frac{H}{a}\int_0^t a(\bar t)(1+q(\bar t)) d{\bar t},
\end{equation}
which implies that if $1+q>0$ then $g(t)>0$ for $t>0$.
A final property that follows from~\eqref{def_g_simple_repeat} is
\begin{equation}
\partial_a(ag)=(1+q)(1-g).  \label{diff_g}
\end{equation}
Thus if $q>-1$ then $ag(a)$ is an increasing function.

Since 1985 the integrals that appear
in the expressions~\eqref{def_g_simple_repeat}
and~\eqref{g_alt} for the function $g$
have appeared in many papers on linear perturbation theory,
usually giving the Bardeen potential $\psi_{\mathrm p}$
for adiabatic long wavelength perturbations.
However, a notation for the function $g$ has not been introduced.
We have already mentioned that Hu and Eisenstein (1999)~\cite{hueis99}
effectively introduced the integral expression for $g(a)$ in this context.
In order to relate our
function $g$ to other work we consider our expression~\eqref{SH_poisson1} for
$\psi_{\mathrm p}$ for adiabatic long wavelength perturbations,
which we write here using $t$ as follows:\footnote
 {Several authors have used the
expression~\eqref{def_g_alt} with $1+q=-{\dot H}/H^2$
for $g(t)$ in~\eqref{psi_p}, \emph{e.g.} Martin and
Schwarz (1998)~\cite{marsch98}, equation (4.26)
and Malik and Wands (2005)~\cite{malwan05}
equation (3.38). }
\begin{equation} \label{psi_p}
\psi_{\mathrm p}(t) \approx Cg(t) - C_{*}\frac{H}{a}=
C\left(1-\frac{H}{a}\int_0^t a(\bar t) d{\bar t}\right) - C_{*}\frac{H}{a}.
\end{equation}
Here $C$ and $C_{*}$ are arbitrary spatial functions. The solution with
$C_{*}=0$ is the growing mode, and is the unique solution
which is bounded as $a\rightarrow 0$. The solution with
$C=0$ is the decaying mode and is unbounded as $a\rightarrow 0$.

If $C\neq0$ then one can incorporate $C_{*}$ into the lower bound of the integral
as follows:
\begin{equation} \label{psi_p,concise}
\psi_{\mathrm p}(t) \approx
C\left(1-\frac{H}{a}\int_{t_{*}}^t a(\bar t) d{\bar t}\right),
\end{equation}
where $t_{*}$ is a spatial function. This is the form in which the expression
for $\psi_{\mathrm p}$ is usually given in the literature. In some
references the expression~\eqref{psi_p,concise} is derived by assuming a
particular matter content, e.g. a perfect fluid with an arbitrary equation of
state (Hwang (1991)~\cite{hwa91a}, see equation (55),
Mukhanov (2005)~\cite{muk05}, see equation (7.69))
or a minimally coupled scalar field ( Mukhanov \emph{et al} (1992)~\cite{muketal92},
see equation (6.56), Hwang (1994)~\cite{hwa94b}, see equation (94)).
It is known, however, that one can derive~\eqref{psi_p}
or~\eqref{psi_p,concise} without
specifying the matter content in detail, as we have done. We refer to
Hu and Eisenstein (1999)~\cite{hueis99}, equation (59),
Bertschinger (2006)~\cite{ber06}, equation (24) with (10) and (11),
noting that his $\kappa$ corresponds to our ${\cal R}$,
and Weinberg (2008)~\cite{wei08}, equations (5.4.16) and (5.4.20).
We note that these authors identify the arbitrary function $C$
in~\eqref{psi_p} with the comoving curvature perturbation ${\cal R}$,
thereby completing the solution.


We  showed in section~\ref{applications} that the function $g$ as
defined by~\eqref{def_g_simple_repeat}
or~\eqref{def_g_alt}, also arises in a perturbed $\Lambda CDM$ cosmology,
in which case it describes the perturbations exactly and on all scales.
In this context, however, it was introduced in a completely different way, namely,
by finding the function $D(a)$, called the growth factor, that is the
appropriately normalized growing solution of
the evolution equation for the linear density perturbation:
\begin{equation}
(\partial_{\eta}\!^2 +{\cal H}\partial_\eta -
\sfrac32 \Omega_m{\cal H}^2)\bdelta_{\mathrm v}=0.
\end{equation}
This function has the following integral expression:\footnote
{See Eisenstein (1997)~\cite{eis97},
equations (3) and (4). Note that his $a$ and $H$ correspond to our $a/a_0$
and $H/H_0$.
 The numerical factor $\sfrac52$ was determined by requiring
that $D(a)/a\rightarrow 1$ as $a\rightarrow 0$. This result was first
given by Heath (1977)~\cite{hea77} using unfamiliar notation.
See also Villa and Rampf (2016)~\cite{vilram16}
equations (5.7) and (5.12)-(5.13), where their
$a$ corresponds to our $a/a_0$.
Matsubara (1995)~\cite{mat95} gives a different representation
of $D$, see equations (8) and (10). }
\begin{equation}\label{D_LCDM}
D(a) = \sfrac52 {\cal H}_0^2\,\Omega_{m,0}
\frac{\cal H}{a} \int_0^a \frac{1}{{\cal H}(\bar a)^3}d{\bar a},
\end{equation}
where ${\cal H}^2$ is given by~\eqref{H_lambdaCDM}.
The numerical factor is fixed by the requirement that
\begin{equation} \label{lim_D}
\lim_{a\rightarrow0}\left(\frac{D(a)}{a/a_0}\right) =1.
\end{equation}
We now relate $D(a)$ to $g(a)$. We begin by writing
the general expression~\eqref{def_g_alt} for $g(t)$ in terms of $a$, obtaining:
\begin{equation} \label{def_g_a}
g(a)= \frac{\cal H}{a^2}\int_0^a \frac{{\bar a}(1+q(\bar a))}{{\cal H} (\bar a)} d{\bar a}.
\end{equation}
In a $\Lambda CDM$ universe it follows from~\eqref{1+w,LCDM}
using $1+q=\sfrac32 (1+w)$ that
\begin{equation}
(a/a_0)(1+q)=\sfrac32 {\cal H}_0^2\,\Omega_{m,0}{\cal H}^{-2}. \label{1+q_LCDM}
\end{equation}
We now specialize  the expression~\eqref{def_g_a}
 to the $\Lambda CDM$
universe by substituting~\eqref{1+q_LCDM}. On comparing the result
 with~\eqref{D_LCDM} we obtain
\begin{equation} \label{g_D}
g(a) = \sfrac35\left(\frac{D(a)}{a/a_0}\right).
\end{equation}
In the $\Lambda CDM$ context the function $g$ was first defined in terms of $D$
in this way, \emph{i.e.} $g(a)$ is proportional to $D(a)/a$. The function
$g$ then determines the Bardeen potential according to
$\psi_{\mathrm p}=g(a)\psi_0(x^i)$ where $\psi_0(x^i)$
is an arbitrary spatial function. See, for example
Bartolo {\it et al} (2006)~\cite{baretal06} (in the text following equation (2.3)) and
Villa and Rampf (2016)~\cite{vilram16} (in the text following equation (5.12)).
The factor $\sfrac35$ in~\eqref{g_D} implies that $\psi_0={\cal R}$.
In the above references this factor is omitted, which
implies that $\psi_0= \sfrac35 {\cal R}.$

%
%

\section{Discussion \label{discussion} }

In this paper we have considered scalar perturbations of
flat FL cosmologies up to second order, subject to the assumption that at first order
the vector and tensor modes are zero. The metric perturbations are
described by the spatially gauge fixed variables
$\phi, \psi, {\cal H}B$. The perturbations of the stress-energy
tensor, which is assumed to have zero anisotropic stresses and
zero heat flux, are described by the variables $\bdelta, {\cal H}V, \Gamma$.
The background stress-energy tensor is characterized by the scalars
$w$, $c_s^2$ and the background dynamics by ${\cal H}$, $q$, where $1+q = \frac32(1+w)$.

Within this framework we have given for the first time the general
explicit solution of the governing equations up to
second order for adiabatic perturbations on super-horizon scales
(see equations~\eqref{comov_sh2}).
We showed that in the total matter gauge
the governing equations can be integrated very easily,
leading to a solution that has a remarkably simple form:
the three matter perturbations are zero and of the three metric perturbations,
one is zero, one is constant in time and the remaining one has an increasing
mode and a decreasing mode\footnote{In cosmological perturbation
theory at second order
the decaying mode is usually set to zero. One
exception is Christopherson \emph{et al} (2016)~\cite{chretal16}.}
with time dependence proportional to $1-g(a)$ and ${\cal H}/a^2$, respectively, at
both first and second order. In other words, \emph{the
perturbation evolution function $g(a)$, which
is determined by the background dynamics through equation~\eqref{def_g_simple},
completely determines the evolution of the growing mode up to second order
for adiabatic perturbations on super-horizon scale}.
Going beyond the initial scope of this paper we showed
in addition that the function $g(a)$ for
$\Lambda CDM$ determines the growing mode of perturbations
of these models \emph{on all scales} to second order.

Having derived the solutions using the total matter
gauge we also obtained the solution in the
uniform curvature gauge and the Poisson gauge by using the change of gauge formulas.
There is an increasing complexity in the solution as one progresses to the
uniform curvature gauge and then to the Poisson gauge, with
the decaying mode adding significantly to the complexity. Moreover, in these
gauges the background scalars $w$ (or $q$) and $c_s^2$
also play a role in determining the evolution.

In a subsequent related paper~\cite{uggwai19c} we  consider second
order perturbations of a flat Friedmann-Lema\^{i}tre
universe whose stress-energy content is a single minimally coupled scalar field with an
arbitrary potential. We apply the methods used in this paper
to derive the general solution of the perturbed Einstein equations in
explicit form for this class of models when the perturbations are in the super-horizon
regime. As a by-product we obtain a new conserved quantity for long wavelength
perturbations of a single scalar field at second order.

\begin{appendix}


\section{Governing equations in the uniform curvature gauge \label{evol_ucg}}

On super-horizon scale the governing equations in the uniform
curvature gauge simplify significantly: the source terms
are independent of $B_{\mathrm c}$ and hence the evolution equation for
$B_{\mathrm c}$ decouples from the other equations. This equation will, however,
not be needed in this paper. The remaining equations, assuming
that the perturbations are adiabatic (${}^{(r)}\!{\Gamma} \approx 0$, $r=1,2$),
have the following
form (specialize the equations in UW2~\cite{uggwai18}, section V.B.1):
\begin{subequations}\label{ucg_gov1}
\begin{align}
(1+q)\partial_N((1+q)^{-1} {}^{(1)}\!{\phi_\mathrm{c}}) &\approx 0,
\label{ucg_gov1.1} \\
{\cal H}{}^{(1)}\!{V}_{\mathrm c} &=  -(1+q)^{-1} {}^{(1)}\!{\phi_\mathrm{c}},\label{ucg_gov1.4}\\
{}^{(1)}\!{\bdelta}_{\mathrm c} &\approx 3 {\cal H}{}^{(1)}\!{V}_{\mathrm c}.
\label{ucg_gov1.3}
\end{align}
\end{subequations}
while at second order we obtain:
\begin{subequations} \label{ucg_gov2}
\begin{align}
(1+q)\partial_N((1+q)^{-1} {}^{(2)}\!{\phi_\mathrm{c}}) \approx
 - \sfrac12{\mathbb S}^{\Gamma}_{\mathrm c}, \\
{\cal H}{}^{(2)}\!{V}_{\mathrm c} \approx - (1+q)^{-1}({}^{(2)}\!{\phi_\mathrm{c}} -
\sfrac12 {\mathbb S}^q_{\mathrm c}),  \\
{}^{(2)}\!{\bdelta}_{\mathrm c} \approx 3{\cal H}{}^{(2)}\!{V}_{\mathrm c} +
\sfrac12(1+q)^{-1}
 ({\mathbb S}^{\rho}_{\mathrm c} -3{\mathbb S}^q_{\mathrm c}).
\end{align}
\end{subequations}
The source terms with kernel $\mathbb S_{\mathrm c}$
are given by (see UW2~\cite{uggwai18}, section V.B.1)
\begin{equation}  \label{source_c}
{\mathbb S}_{\mathrm c}={\mathbb G}_{\mathrm c}-3(1+w){\mathbb T}_{\mathrm c},
\end{equation}
where the Einstein tensor source terms are
\begin{equation}
{\mathbb G}^{\Gamma}_{\mathrm c} \approx
-8{\cal L}_1\phi_{\mathrm c}^2=
-8(1+q)\partial_N\left((1+q)^{-1} \phi_\mathrm{c}^2\right) , \quad
{\mathbb G}^q_{\mathrm c} \approx   8\phi_{\mathrm c}^2 ,\quad
{\mathbb G}^{\rho}_{\mathrm c} \approx 24\phi_{\mathrm c}^2 ,
\end{equation}
(see equation (34a) in UW2~\cite{uggwai18} for
the definition of the differential operator ${\cal L}_1$)
and the stress-energy source terms are
\begin{subequations}  \label{source.T_c}
\begin{align}
{\mathbb T}^{\Gamma}_{\mathrm c} &\approx
 - \sfrac13(\partial_N c_s^2)\bdelta_{\mathrm c}^2,  \\
{\mathbb T}^q_{\mathrm c} &\approx 2{\cal S}^i\left[\left((1+c_s^2)\bdelta_{\mathrm c} - \phi_{\mathrm c}\right){\bf D}_i({\cal H}{V_{\mathrm c}})\right], \\
{\mathbb T}^{\rho}_{\mathrm c} &\approx 0.
\end{align}
\end{subequations}
Here and elsewhere in this Appendix, in order to simplify
the notation we have dropped the superscript ${}^{(1)}$ on the
linear perturbations in the source terms.
\section{The density perturbation constraint}\label{app:density}

We restrict the general expression for the density perturbations
${}^{(r)}\!\bdelta,\,r=1,2$, valid in any temporal gauge,
given in UW2~\cite{uggwai18} (see equations (40)) to super-horizon scales:
\begin{subequations} \label{delta_gen}
\begin{align}
{}^{(1)}\!\bdelta &\approx 3{\cal H}{}^{(1)}\!V ,  \label{delta1_gen}\\
{}^{(2)}\!\bdelta &\approx3{\cal H}{}^{(2)}\!V +
{\mathbb S}^{\rho} - 3{\mathbb S}^q, \label{delta2_gen}
\end{align}
where
\begin{equation}
{\mathbb S}^{\rho} = {\mathbb G}^{\rho} - 3(1+w){\mathbb T}^{\rho}, \qquad
{\mathbb S}^q = {\mathbb G}^q - 3(1+w){\mathbb T}^q.  \label{delta2_source}
\end{equation}
\end{subequations}
On specializing the source terms ${\mathbb G}$ and ${\mathbb T}$
to super-horizon scales and using
the equation $(1+q){\cal H}{}^{(1)}\!V = -(\partial_{N}{}^{(1)}\!\psi +{}^{(1)}\!\phi)$
we obtain
\begin{equation}
{\mathbb S}^{\rho} - 3{\mathbb S}^q \approx 3(1+q)({\cal H}V)^2 +
(1+c_s^2){\bdelta}^2 + 6{\cal S}^i(\Gamma {\bf D}_i ({\cal H}V)).
\end{equation}
On introducing the hatted variables as defined
by equations~\eqref{hatted_var}, equation~\eqref{delta2_gen}
assumes the concise form
\begin{equation} \label{delta2_gen_SH}
{}^{(2)}\!{\hat\bdelta} \approx 3{\cal H}{}^{(2)}\!{\hat V} +
6{\cal S}^i(\Gamma {\bf D}_i ({\cal H}V)),
\end{equation}
valid for any temporal gauge. The scalar mode extraction operator ${\cal S}^i$ is given
by ${\cal S}^i = {\bf D}^{-2}{\bf D}^i$, where ${\bf D}^{-2}$ is the inverse spatial Laplacian.

\section{Change of gauge formulas\label{app:changegauge}}

We require the following change of gauge formulas
for long wavelength perturbations that can be obtained from
UW1~\cite{uggwai19a} (specialize the formulas at
 the end of section 3 by dropping terms of order two or higher
 in ${\bf D}_i$):
\begin{subequations}\label{boxmaster}
\begin{align}
{}^{(2)}\!\hat{\Box}_\mathrm{v} &= {}^{(2)}\!\hat{\Box} -
{\cal H}{}^{(2)}\!\hat {V} + 2(\partial_N \Box_\mathrm{v}){\cal H}V  +
\Box_{rem,\mathrm{v}} +2{\cal S}^i [\phi_{\mathrm v}({\bf  D}^i {\cal H}V)],
 \label{box_change_v2} \\
{}^{(2)}\!\hat{\Box}_{\mathrm p} &= {}^{(2)}\!\hat{\Box} - {\cal H}{}^{(2)}\!\hat {B} +
2(\partial_N \Box_{\mathrm p}){\cal H}B +
\Box_{rem,{\mathrm p}} - {\cal H}B_{rem, {\mathrm p}}, \label{box_change_v1}
\end{align}
\end{subequations}
where the kernel $\Box$ can be one of $\psi, {\cal H}B,
{\cal H}V$ or $\sfrac13{\bdelta}$, and the gauge on the right
side can be one of the standard choices.
 Equation~\eqref{box_change_v2} can be specialized to give the following generalizations of
some of the first order gauge formulas:\footnote
{Choose $\Box=\psi$ with first the uniform curvature gauge on the right side
and then the Poisson gauge and use~\eqref{totmat_1.2new}
($\partial_N {\psi}_{\mathrm v}
 = -\phi_{\mathrm v}$).
Then choose $\Box=\sfrac13{\bdelta}$ with the Poisson gauge
on the right side and use~\eqref{delta_v_zero} ($\!\bdelta_{\mathrm v}\approx 0$).}
\begin{subequations}  \label{gc_comov}
\begin{align}
{}^{(2)}\!{\hat\psi}_{\mathrm v}&\approx  -
{\cal H}{}^{(2)}\!{\hat V}_{\mathrm c}-
2{\cal S}^i [({\bf  D}^i\phi_{\mathrm v}) {\cal H}V_{\mathrm c}], \label{change_uc1} \\
{}^{(2)}\!{\hat\psi}_{\mathrm v}&\approx {}^{(2)}\!{\hat\psi}_{\mathrm p} -
{\cal H}{}^{(2)}\!{\hat V}_{\mathrm p} -
2{\cal S}^i [({\bf  D}^i\phi_{\mathrm v}) {\cal H}V_{\mathrm p}],   \label{poisson2} \\
 {}^{(2)}\!{\hat \bdelta}_{\mathrm p} &\approx
3{\cal H}{}^{(2)}\!{\hat V}_{\mathrm p} -
6{\cal S}^i [\phi_{\mathrm v}({\bf  D}^i {\cal H}V_{\mathrm p})].   \label{poisson3}
\end{align}
\end{subequations}
These formulas simplify further and match the corresponding first order formulas if
the perturbations are also adiabatic and the Einstein equations hold since then
${}^{(1)}\!\phi_{\mathrm v}\approx 0.$

Next choose $\Box={\cal H}B$ in~\eqref{box_change_v2}
with the uniform curvature gauge on the right side.
On using~\eqref{change_uc1}, the relation ${}^{(1)}\!\psi_{\mathrm v}=  -
{\cal H}{}^{(1)}\! V_{\mathrm c}$
 and the first order solution~\eqref{totmat.soln1}
we obtain the
following more complicated relation:
\begin{subequations}\label{gc_Bv2}
\begin{equation}
{\cal H}{}^{(2)}\!\hat{B}_\mathrm{c} \approx {\cal H}{}^{(2)}\!\hat{B}_{\mathrm v} -
{}^{(2)}\!{\hat\psi}_{\mathrm v} +
2\partial_N ({\cal H}B_\mathrm{v})\psi_{\mathrm v}  -
{\cal H}B_{rem,\mathrm{v},\mathrm{c}},
\end{equation}
where
\begin{equation}
\begin{split}
{\cal H}B_{rem,\mathrm{v},\mathrm{c}} &\approx
(\partial_{N} +2q)\left({\mathbb D}_0({\cal H}B_\mathrm{v})-
{\mathbb D}_0({\cal H}B_{\mathrm c})\right) \\& \qquad
+ 2{\cal S}^i\left[(\phi_\mathrm{v} +\phi_{\mathrm p}){\bf D}_i{\cal H}B_\mathrm{v}  -
(\phi_{\mathrm c} +\phi_{\mathrm p}){\bf D}_i{\cal H}B_{\mathrm c} \right]. \label{rem_B}
\end{split}
\end{equation}
\end{subequations}
Next choose $\Box=\psi$ in~\eqref{box_change_v1} and use the total matter gauge
on the right side to obtain:
\begin{equation}
\begin{split}
{}^{(2)}\!{\hat\psi}_{\mathrm p}\approx {}^{(2)}\!{\hat\psi}_{\mathrm v} -
{\cal H}{}^{(2)}\!{\hat B}_{\mathrm v}& +
2(\partial_{N}\psi_{\mathrm p}){\cal H}B_{\mathrm v} +
(\partial_{N}+2q){\mathbb D}_0({\cal H}B_{\mathrm v})  \\&+
2{\cal S}^i[(\phi_{\mathrm p} +\phi_{\mathrm v}){\bf D}_i({\cal H}B_{\mathrm v})].
\end{split}
 \label{poisson1}
\end{equation}
In addition the perturbed Einstein equations in the
Poisson gauge UW2~\cite{uggwai18} (introduce hatted
variables in equation (48b) in~\cite{uggwai18})  yield:
\begin{equation}
{}^{(2)}\!{\hat\phi}_{\mathrm p}\approx {}^{(2)}\!{\hat\psi}_{\mathrm p} -
4[{\mathbb D}_0(\psi_{\mathrm p}) +
(1+q){\mathbb D}_0({\cal H}V_{\mathrm p})]. \label{poisson4}
\end{equation}
The source terms in equations~\eqref{gc_Bv2} and~\eqref{poisson1}
can be evaluated using the first order solutions in
sections~\ref{totmat_soln}-~\ref{Poisson_soln},
and the derivative $\partial_{N}g= (1+q)(1-g) -g$
which follows from~\eqref{diff_g}.

Finally we show that the uniform density gauge is equivalent to the total matter
gauge on super-horizon scales to second order. This is a consequence
of the relations~\eqref{delta_v_zero},~\eqref{delta1_gen}
and~\eqref{delta2_gen_SH}, which
imply that ${}^{(r)}\!V_{\rho}=0,\,r=1,2$,
and the fact that the metric perturbations $f=(\phi,\psi,B)$
satisfy ${}^{(r)}\!{f}_{\rho}\approx{}^{(r)}\!{f}_{\mathrm v}$
for $r=1,2$, on super-horizon scales
when the perturbed Einstein equations at linear order hold, where
the latter result follows from UW1~\cite{uggwai19a}.\footnote
{Choose the total matter gauge in equation (41d) which yields
$\xi^N_{\rho,\mathrm v}\approx0$, and then use (39a,b) and (40b).}

\end{appendix}

\bibliographystyle{plain}
\bibliography{../Bibtex/cos_pert_papers}

\end{document}